\documentstyle[amstex,amssymb,epsfig,12pt]{article}

\input{tcilatex}

\begin{document}

\bigskip 
\begin{titlepage}  
\bigskip \begin{flushright}  
\end{flushright}

\vspace{1cm}  
  
\begin{center}  
{\Large \bf {Entropy and Mass Bounds of Kerr-de Sitter Spacetimes}}\\
\end{center}
\vspace{2cm}
\begin{center}
 A.M. Ghezelbash{%
\footnote{%
EMail: amasoud@@sciborg.uwaterloo.ca}} and R. B. Mann\footnote{
EMail: mann@@avatar.uwaterloo.ca}\\
Department of Physics, University of Waterloo, \\
Waterloo, Ontario N2L 3G1, Canada\\
\vspace{1cm}
\end{center}

\begin{abstract}
\end{abstract}
We consider Kerr-de Sitter spacetimes and evaluate their mass, angular momentum and entropy according
to the boundary counterterm prescription. We provide a physicall interpretation for angular velocity
and angular momentum at future/past infinity. We show that the entropy of the 
four-dimensional Kerr-de Sitter spacetimes is less than of pure de Sitter spacetime 
in agreement to the entropic N-bound. Moreover, we show that maximal mass conjecture
which states {\it ``any asymptotically de Sitter spacetime with mass greater than de Sitter has a
cosmological singularity"} is respected by asymptotically de Sitter spacetimes with rotation. 
We furthermore consider the possibility of strengthening the conjecture to state 
that {\it ``any asymptotically dS spacetime will have mass 
greater than dS if and only if it has a cosmological singularity''} and find that 
Kerr-de Sitter spacetimes do not respect this stronger statement.
We investigate the behavior of the $c$-function for the  Kerr-de Sitter spacetimes and
show that it is no longer isotropic. However an average of 
the $c$-function over the angular variables yields a renormalization group flow 
in agreement with the expansion of spacetime at future infinity.

\end{titlepage}\onecolumn

\begin{center}
\bigskip 
\end{center}

\section{Introduction}

There is still much to be learned about holographic duality, particularly
for spacetimes that are asymptotically de Sitter (dS). Amongst the many
challenges presented in establishing (or perhaps refuting) a dS holographic
correspondence principle is the calculation of conserved quantities. Unlike
their asymptotically flat or asymptotically anti de Sitter counterparts,
asymptotically de Sitter spacetimes have neither a spatial infinity nor a
global timelike Killing vector. Consequently both the definition and the
computation of conserved charges for such spacetimes are not straightforward.

However it has recently been shown that it is possible under certain
circumstances \ to compute such charges at past or future infinity \cite%
{bala}. This method, analogous to the Brown-York prescription in
asymptotically AdS spacetimes \cite{balakraus,brown,BCM,ivan}, is inspired
from a conjectured holographic duality between the physics in asymptotically
dS spacetimes and that of a boundary Euclidean conformal field theory (CFT)
that resides on past or future infinity. The specific prescription in ref. %
\cite{bala} (employed previously by others in more restricted contexts \cite%
{Klem, Myung}) can be generalized to an arbitrary number of dimensions, in
which there is an algorithmic prescription for adding boundary terms that
render the action finite \cite{GM, Dumitru}.

Carrying out a procedure analogous to that in the AdS case \cite%
{balakraus,BCM}, it is straightforward to compute the boundary stress
tensor, and from this obtain conserved charges associated with the
asymptotically dS (adS) spacetime at future/past infinity. In particular the
conserved charge associated with the (asymptotic) Killing vector $\partial
/\partial t$ -- now spacelike outside of the cosmological horizon -- can be
interpreted as the conserved mass. \ Employing this definition, the authors
of ref. \cite{bala} were led to the conjecture that {\it any asymptotically
dS spacetime with mass greater than dS has a cosmological singularity. }\ We
shall refer to this conjecture as the maximal mass conjecture. As stated the
conjecture is in need of clarification before a proof can be considered, but
roughly speaking it means that the conserved mass of any physically
reasonable adS spacetime must be negative (i.e. less than the zero value of
pure dS spacetime). \ This has been verified for topological dS solutions
and their dilatonic variants \cite{cai} and for Schwarzschild-de Sitter
(SdS) black holes up to dimension nine \cite{GM}. The maximal mass
conjecture was based in part on the Bousso N-bound \cite{bousso}, another
conjecture stating that {\it any asymptotically dS spacetime will have an
entropy no greater than the entropy }$\pi \ell ^{2}${\it \ of pure dS with
cosmological constant }$\Lambda =3/\ell ^{2}${\it \ in }$(3+1)${\it \
dimensions}.

Recently it has been shown that locally asymptotically dS spacetimes with
NUT charge furnish counterexamples to both of these conjectures \cite%
{CGM1,CGM2}. \ Specifically, there is a range of parameter space in which
the conserved mass of the NUT-charged dS spacetime can be greater than zero,
and/or its entropy exceeds that of pure dS spacetime. Note that outside of
the cosmological horizon NUT-charged dS spacetimes are not afflicted with
closed timelike curves the way that their asymptotically flat and anti de
Sitter counterparts are.

In this paper we extend these considerations to charged dS spacetimes with
rotation. There are several reasons for being interested in this class of
spacetimes. One is to understand the physical interpretation of angular
momentum outside of the horizon as well as the role and the contribution of
the rotation to the entropy. Another is to understand the nature of the $c$%
-function \cite{stro} when rotation is present. \ We find that in general it
becomes a function of angle, and we describe how to physically interpret it
in this more general setting. Yet another reason is to explore the validity
and utility of the maximal mass conjecture when rotation is present. \ One
might hope that the maximal mass conjecture, if correct, might serve as a
diagnostic tool in determining the presence/absence of singularities. \ This
would be the case if the conjecture could be strengthened to state that {\it %
any asymptotically dS spacetime will have mass greater than dS if and only
if it has a cosmological singularity}. This stronger version of the
conjecture would then imply that a spacetime with negative mass at future
infinity would be free of cosmological singularities. \ 

To this end we shall consider in this paper spacetimes with zero NUT charge,
deferring the investigation of rotating asymptotically dS spacetimes with
non-vanishing NUT charge to future work \cite{GM2}. We proceed as follows.
We consider the Kerr-dS spacetime and review the procedure for calculating
the conserved mass, angular momentum and entropy of the spacetime. We then
examine the behavior of different physical quantities of the spacetime and
provide a physical interpretation of quantities such as angular velocity and
angular momentum at future (past) infinity. We find that the N-bound and
maximal mass conjectures are never violated even when the black hole is
extremal. However we also find that Kerr-dS spacetime can have naked
singularities even when the mass (at future infinity) is negative.
Consequently it is difficult to see how the stronger version of the
conjecture can be realized for any spacetimes of physical interest. We then
extend the notion of the $c$-function \cite{stro} to this situation and show
that it depends on the angular coordinate, due to rotation. We then average
the $c$-function over the angular variables and show that the
renormalization group flow is in agreement with the expansion of spacetime
at future infinity.

\bigskip

\section{Kerr-dS spacetime}

\bigskip \label{Kerr-dS sec}

The Euclidean Kerr-dS geometry is given by the line element%
\begin{equation}
ds^{2}=\frac{\Delta _{E}(r)}{\Xi _{E}^{2}\rho _{E}^{2}}(dt+a\sin ^{2}\theta
d\phi )^{2}+\frac{\Theta _{E}(\theta )\sin ^{2}\theta }{\Xi _{E}^{2}\rho
_{E}^{2}}[adt-(r^{2}-a^{2})d\phi ]^{2}+\frac{\rho _{E}^{2}dr^{2}}{\Delta
_{E}(r)}+\frac{\rho _{E}^{2}d\theta ^{2}}{\Theta _{E}(\theta )}
\label{KdSEmetr}
\end{equation}%
where 
\begin{equation}
\begin{array}{c}
\rho _{E}^{2}=r^{2}-a^{2}\cos ^{2}\theta \\ 
\Delta _{E}(r)=-\frac{r^{2}(r^{2}-a^{2})}{\ell ^{2}}+r^{2}-2mr-a^{2} \\ 
\Theta _{E}(\theta )=1-\frac{a^{2}}{\ell ^{2}}\cos ^{2}\theta \\ 
\Xi _{E}=1-\frac{a^{2}}{\ell ^{2}}%
\end{array}
\label{KdSEfuncs}
\end{equation}%
For \ $a\leq \ell ,$ the metric function $\Theta _{E}(\theta )$ always is
positive and the Euclidean section exists for some values of the coordinate $%
r$ between different successive roots of $\Delta _{E}(r)$, such that $\Delta
_{E}(r)$ is a positive valued function there. Depending on the values of the
rotation parameter $a,$ the cosmological parameter $\ell =\sqrt{3/\Lambda }$
(where $\Lambda $ \ is the cosmological constant) and the mass parameter $m$%
, the function $\Delta _{E}(r)$ can have between two to four real roots.

The Lorentzian geometry is given by%
\begin{equation}
ds^{2}=-\frac{\Delta _{L}(r)}{\Xi _{L}^{2}\rho _{L}^{2}}(dt-a\sin ^{2}\theta
d\phi )^{2}+\frac{\Theta _{L}(\theta )\sin ^{2}\theta }{\Xi _{L}^{2}\rho
_{L}^{2}}[adt-(r^{2}+a^{2})d\phi ]^{2}+\frac{\rho _{L}^{2}dr^{2}}{\Delta
_{L}(r)}+\frac{\rho _{L}^{2}d\theta ^{2}}{\Theta _{L}(\theta )}
\label{KdSLmetr}
\end{equation}

\bigskip where%
\begin{equation}
\begin{array}{c}
\rho _{L}^{2}=r^{2}+a^{2}\cos ^{2}\theta \\ 
\Delta _{L}(r)=-\frac{r^{2}(r^{2}+a^{2})}{\ell ^{2}}+r^{2}-2mr+a^{2} \\ 
\Theta _{L}(\theta )=1+\frac{a^{2}}{\ell ^{2}}\cos ^{2}\theta \\ 
\Xi _{L}=1+\frac{a^{2}}{\ell ^{2}}%
\end{array}
\label{KdSLfuncs}
\end{equation}%
The event horizons of the spacetime are given by the singularities of the
metric function, which are the real roots of $\Delta _{L}(r)=0$. Hence the
horizons are determined by the solutions of the equation%
\begin{equation}
r_{H}^{4}-r_{H}^{2}(\ell ^{2}-a^{2})+2m\ell ^{2}r_{H}-\ell ^{2}a^{2}=0
\label{horizons}
\end{equation}%
Note that the roots of this equation are not the same as those of the
function $\Delta _{E}(r)$. The smallest root of $\Delta _{L}(r)$ is
negative, while the other three roots corresponding to inner, outer and
cosmological horizons are positive. The Lorentzian metric function $\Delta
_{L}(r)\geq 0$ between its smallest root and the inner horizon and between
the outer and cosmological horizons, where $t$ is a timelike coordinate. It
is negative elsewhere, where $t$ is spacelike.

In the limit $\ell \rightarrow \infty ,$ equation (\ref{horizons}) yields
the well known location of the Kerr black hole horizon 
\begin{equation}
r_{H}=m+\sqrt{m^{2}-a^{2}}  \label{Kerrhorizon}
\end{equation}%
When the rotational parameter is very small $a\rightarrow 0,$ then the
equation (\ref{horizons}) reduces to 
\begin{equation}
\frac{r_{H}^{3}}{\ell ^{2}}-r_{H}+2m=0  \label{horizonsSdS}
\end{equation}%
which gives us the location of Schwarzschild-dS event horizon $r_{H}$ and
cosmological horizon $r_{C}$. In this case, for mass parameters $m$ with $%
0<m<m_{N}$, where 
\begin{equation}
m_{N}=\frac{\ell }{3\sqrt{3}}  \label{naraimass}
\end{equation}%
we have a black hole in dS spacetime with event horizon at $r=r_{H}$ and
cosmological horizon at $r=r_{C}>r_{H}.$ When $m=m_{N},$ the event horizon
coincides with the cosmological horizon $r_{C}=r_{H}$ $=\frac{\ell }{\sqrt{3}%
}$ and one gets the rotating Nariai solution. For $m>m_{N},$ the
Schwarzschild-dS metric describes a naked singularity in an asymptotically
dS spacetime. So demanding the absence of naked singularities yields an
upper limit to the mass of the Schwarzschild-dS black hole. The other
extreme case is when $a\rightarrow \infty $\ (i.e. negligible $m$), which
can straightforwardly be shown to be pure dS. Eq. (\ref{horizons}) gives%
\begin{equation}
r_{H}=\ell  \label{horizonsainf}
\end{equation}%
for the horizon.

In general for the metric (\ref{KdSLmetr}), the rotating Nariai solution has
an event horizon (coincident with its cosmological horizon) at 
\begin{equation}
r_{H}=\frac{3m+\sqrt{9m^{2}-8a^{2}(1-\frac{a^{2}}{\ell ^{2}})}}{2(1-\frac{%
a^{2}}{\ell ^{2}})}  \label{rHNaraiKdS}
\end{equation}%
which reduces to $r_{H}$ $=\frac{\ell }{\sqrt{3}}$ when $a=0.$ For a fixed
value of $a$, we can find from (\ref{horizons}) the following equations for
the extremum of the horizon radius%
\begin{equation}
2r_{H}+m\mp \sqrt{m^{2}+8mr_{H}}-2r_{H}^{3}/\ell ^{2}=0  \label{exthor}
\end{equation}%
for which the upper branch has a maximum at $r_{+}=(\frac{3}{4}+\frac{\sqrt{3%
}}{2})m$ for fixed $\ell $. The corresponding critical cosmological
parameter is $\ell _{+}=\frac{m}{4}(2\sqrt{3}+3)^{3/2}$. In this case from
eq. (\ref{horizons}), we find that $r_{+}$\ is a triple root and the
corresponding rotational parameter is $a=\frac{m}{4}\sqrt{6\sqrt{3}+9}$.
This case is a special case of general rotating Nariai solution where all
three horizons coincide to each other.

Other interesting special cases of the Kerr-dS metric are the cold metrics
(where the inner and outer horizons coincide) and the lukewarm metrics
(where the surface gravities of the cosmological and outer horizons are the
same). \ A full analysis of these cases has been carried out previously \cite%
{ivanrobb} and we shall not reproduce these results here.

{\large \ \ }Denoting the four real roots (which collectively sum to zero)
of $\Delta _{L\text{ }}(r)$ in increasing order by $-c-\alpha ,-c+\alpha
,c-\beta ,c+\beta ,$ where $0\leq \beta <c$ and $c<\alpha \leq 2c-\beta ,$
we see that the first root is negative and that the other three roots
corresponding to inner, outer and cosmological horizons are positive. In the
limit $\beta \rightarrow 0,$ we have a rotating Nariai solution \cite%
{ivanrobb} for which the metric becomes 
\begin{equation}
ds_{Nariai}^{2}=-\widetilde{\Delta }_{L}(\widetilde{r})\widetilde{\rho }%
_{L}^{2}d\tau ^{2}+\frac{\Theta _{L}(\theta )\sin ^{2}\theta }{\widetilde{%
\rho }_{L}^{2}}[2ac\widetilde{r}d\tau +\frac{c^{2}+a^{2}}{\Xi _{L}}d\varphi
]^{2}+\widetilde{\rho }_{L}^{2}(\frac{d\widetilde{r}^{2}}{\widetilde{\Delta }%
_{L}(\widetilde{r})}+\frac{d\theta ^{2}}{\Theta _{L}(\theta )})
\label{KdSLnariaimetr}
\end{equation}%
where the new coordinates $\widetilde{r},\varphi $ and $\tau $\ are given by%
\begin{equation}
\begin{array}{c}
\widetilde{r}=\frac{r-c}{\beta } \\ 
\varphi =\phi -\frac{a}{a^{2}+c^{2}}t \\ 
\tau =\frac{\beta }{(a^{2}+c^{2})\Xi _{L}}t%
\end{array}
\label{KdSLcoords}
\end{equation}%
\bigskip and $\widetilde{\Delta }_{L}(\widetilde{r})=\frac{1}{\ell ^{2}}%
(2c-\alpha )(2c+\alpha )(1-\widetilde{r}^{2}),$ $\widetilde{\rho }%
_{L}^{2}=c^{2}+a^{2}\cos ^{2}\theta .$ Note that the metric (\ref%
{KdSLnariaimetr}) is a special case (i.e. no electric and magnetic charges)
of the rotating Nariai solution for Kerr-dS spacetimes with electric and
magnetic charges \cite{ivanrobb}. In the general case, the solution is given
by the metric (\ref{KdSLnariaimetr}) and an electromagnetic potential. In
the limiting case $a\rightarrow 0${\bf ,} the solution reduces to the
non-rotating charged Nariai solution considered in \cite{simon}.

The Nariai solution is in thermal equilibrium due to the coincidence of the
outer and cosmological horizons, with common temperature $T_{Nariai}=\frac{1%
}{4\pi \ell ^{2}}(4c^{2}-\alpha ^{2}).$ In general, the non-extremal
(electric and/or magnetic charged) Kerr-dS black holes are not in thermal
equilibrium since the temperatures of the outer and cosmological horizons
are not the same. The exceptions to this rule are the Nariai solution and
the lukewarm solution, that is the solution in which the cosmological and
outer black hole horizons remain distinct, but have identical surface
gravities. This occurs when the rotational parameter takes on the particular
values given by the relation $a^{2}=c^{2}-\frac{\alpha ^{2}+\beta ^{2}}{2}$ %
\cite{ivanrobb}.

Outside of the cosmological horizon, the Kerr-dS metric function $\Delta
_{L}(r)$\ is negative, so we set $r=\tau $ and rewrite the line element in
the form%
\begin{equation}
ds^{2}=-\frac{\rho ^{2}d\tau ^{2}}{\Delta (\tau )}+\frac{\Delta (\tau )}{\Xi
^{2}\rho ^{2}}(dt-a\sin ^{2}\theta d\phi )^{2}+\frac{\Theta (\theta )\sin
^{2}\theta }{\Xi ^{2}\rho ^{2}}[adt-(\tau ^{2}+a^{2})d\phi ]^{2}+\frac{\rho
^{2}d\theta ^{2}}{\Theta (\theta )}  \label{KdSoutmetr}
\end{equation}%
where%
\begin{equation}
\begin{array}{c}
\rho ^{2}=\tau ^{2}+a^{2}\cos ^{2}\theta \\ 
\Delta (\tau )=\frac{\tau ^{2}(\tau ^{2}+a^{2})}{\ell ^{2}}-\tau ^{2}+2m\tau
-a^{2} \\ 
\Theta (\theta )=\Theta _{L}(\theta ) \\ 
\Xi =\Xi _{L}%
\end{array}
\label{KdSoutfuncs}
\end{equation}%
The angular velocity of the horizon is given by%
\begin{equation}
\Omega _{H}=\left. -\frac{g_{t\phi }}{g_{\phi \phi }}\right| _{\tau =\tau
_{c}}=\frac{a}{\tau _{c}^{2}+a^{2}}  \label{KdSangvel}
\end{equation}%
where $\tau _{c}$ is the cosmological horizon $(\Delta (\tau _{c})=0$ and $%
\Delta (\tau >\tau _{c})>0)$. The Killing vector $\chi ^{\mu }=\zeta ^{\mu
}+\Omega _{H}\psi ^{\mu \text{ }}$ is normal to the cosmological horizon $%
\tau =\tau _{c}$, where $\zeta ^{\mu }=(0,1,0,0)$ is the stationary Killing
vector and $\psi ^{\mu \text{ }}=(0,0,0,1)$ is the axial Killing vector with
respect to coordinate system $(\tau ,t,\theta ,\phi ).$ The surface gravity
of the black hole on the horizon is given by $\kappa =\sqrt{\frac{-1}{2}%
\bigtriangledown ^{\mu }\chi ^{\nu }\bigtriangledown _{\mu }\chi _{\nu }}$ %
\cite{WALD} which becomes%
\begin{equation}
\kappa =\frac{1}{2(\tau _{c}^{2}+a^{2})\Xi }\left. \frac{d\Delta }{d\tau }%
\right| _{\tau =\tau _{c}}  \label{surfgra}
\end{equation}%
To compute the conserved mass and the entropy of the spacetime outside the
cosmological horizon, we consider the four dimensional action that yields
the Einstein equations with a positive cosmological constant 
\begin{equation}
I=\frac{1}{16\pi }\int_{{\cal M}}d^{4}x\sqrt{-g}\left( R-2\Lambda \right) -%
\frac{1}{8\pi }\int_{{\cal \partial M}^{-}}^{{\cal \partial M}^{+}}d^{3}x%
\sqrt{h^{\pm }}K^{\pm }+I_{ct}  \label{totaction}
\end{equation}%
where ${\cal \partial M}^{\pm }$ are the future/past boundaries, and $\int_{%
{\cal \partial M}^{-}}^{{\cal \partial M}^{+}}d^{3}x$ indicates an integral
over a future boundary minus an integral over a past boundary, with
respective induced metrics $h_{\mu \nu }^{\pm }$ and intrinsic/extrinsic
curvatures $K_{\mu \nu }^{\pm }$ and $\hat{R}\left( h^{\pm }\right) $
induced from the bulk spacetime metric $g_{\mu \nu }$. $I_{ct}$ is the
counter-term action, calculated to cancel the divergences from the first two
terms (given in \cite{GM}) and we set the gravitational constant $G=1$. The
associated boundary stress-energy tensor is obtained by the variation of the
action with respect to the boundary metric, the explicit form of which can
be found in \cite{GM}.

If the boundary geometries have an isometry generated by a Killing vector $%
\xi ^{\pm }$, then it is straightforward to show that $T_{ab}^{\pm }\xi
^{\pm b}$ is divergenceless, from which it follows that there will be a
conserved charge ${\frak Q}^{\pm }$ between surfaces of constant $t$, whose
unit normal is given by $n^{\pm a}$. Physically this means that a collection
of observers on the hypersurface all observe the same value of ${\frak Q}$
provided this surface had an isometry generated by $\xi ^{b}$ (for explicit
calculations, see{\it \ }\cite{GM}). If $\partial /\partial t$ is itself a
Killing vector, then this can be defined as ${\frak Q}^{\pm }={\frak M}^{\pm
}$, the conserved mass associated with the future/past surface $\Sigma ^{\pm
}\left( \tau \right) $ at any given point $t$ on the boundary. This quantity
changes with the cosmological time $\tau $. Since all asymptotically dS
spacetimes must have an asymptotic isometry generated by $\partial /\partial
t$, there is at least the notion of a conserved total mass ${\frak M}^{\pm }$
for the spacetime in the limit that $\Sigma ^{\pm }$ are future/past
infinity.

\section{Action, Mass and Entropy}

For the Killing vector $\zeta ^{\mu }$, which is spacelike outside the
cosmological horizon, we obtain the conserved mass%
\begin{equation}
{\frak M}=-\frac{m}{\Xi ^{2}}+\frac{4a^{4}+5\ell ^{4}}{40\ell ^{2}\Xi ^{2}}%
\frac{1}{\tau }+O(\frac{1}{\tau ^{2}})  \label{totmass}
\end{equation}%
for the metric (\ref{KdSoutmetr}) near future infinity. This result holds
even for the charged Kerr-dS spacetimes \cite{Deh}.

We note that for $a=0$ (and $\Xi =1$), the total mass (\ref{totmass})
reduces exactly to the total mass of the four dimensional Schwarzschild-dS
black hole \cite{GM}. We observe that for all the Kerr-dS black holes with
positive mass parameter, the total mass is negative, satisfying the maximal
mass conjecture \cite{bala} since the spacetime (\ref{KdSoutmetr}) is free
of any cosmological singularities. The total action (\ref{totaction}) of the
spacetime is 
\begin{equation}
I=-\frac{\beta _{H}(\tau _{c}^{3}+a^{2}\tau _{c}+m\ell ^{2})}{2\ell ^{2}\Xi
^{2}}+\frac{\beta _{H}(4a^{4}+20a^{2}\ell ^{2}+15\ell ^{4})}{120\ell ^{2}\Xi
^{2}}\frac{1}{\tau }+O(\frac{1}{\tau ^{2}})  \label{acttot}
\end{equation}%
where $\beta _{H}$ is the analogue of the Hawking temperature outside of the
cosmological horizon. It is related to the surface gravity of the horizon $%
\kappa $ by%
\begin{equation}
\beta _{H}=\left| \frac{2\pi }{\kappa }\right| =\frac{4\pi (\tau
_{c}^{2}+a^{2})\Xi }{\left| \left. \frac{d\Delta }{d\tau }\right| _{\tau
=\tau _{c}}\right| }=\frac{2\pi (\tau _{c}^{2}+a^{2})(\ell ^{2}+a^{2})}{%
\left| 2\tau _{c}^{3}+\tau _{c}(a^{2}-\ell ^{2})+m\ell ^{2}\right| }
\label{betaH}
\end{equation}%
We note that in the limit $a\rightarrow 0,$ the above $\beta _{H}$
approaches $\frac{2\pi \tau _{c}^{2}\ell ^{2}}{2\tau _{c}^{3}-\tau _{c}\ell
^{2}+m\ell ^{2}}=\frac{2\pi \tau _{c}^{2}\ell ^{2}}{\tau _{c}^{3}-m\ell ^{2}}
$ where we use the relation $\tau _{c}\ell ^{2}=\tau _{c}^{3}+2m\ell ^{2}$
for the location of Schwarzschild-dS cosmological horizon, to eliminate the
linear term of $\tau _{c}$ in the denominator. The last quantity is in exact
agreement with the analogue of the Hawking temperature outside of the
cosmological horizon of the Schwarzschild-dS black hole \cite{GM}. The other
conserved charge, associated with axial Killing vector $\psi ^{\mu
}=(0,0,0,1)$ is the angular momentum%
\begin{equation}
{\frak J}=-\frac{am}{\Xi ^{2}}  \label{totangumom}
\end{equation}%
which in the limiting case of Schwarzschild-dS $(a=0)$ is zero.

Extending the definition of entropy to asymptotically dS spacetimes \cite%
{GM,CGM2} with rotation, we employ the relation $S=\lim_{\tau \rightarrow
\infty }\left( \beta _{H}{\cal M}-I\ \right) $ where ${\cal M}$ is the
conserved charge associated with the Killing vector $\chi ^{\mu }$. Note
that ${\cal M}$ is distinct from the mass ${\frak M}$ in eq. (\ref{totmass}%
). We find for the entropy 
\begin{equation}
S=\frac{\pi (\tau _{c}^{2}+a^{2})}{\Xi }  \label{KdSoutent2}
\end{equation}%
which is exactly $1/4$\ of the area of the cosmological horizon. We note
that the following relations%
\begin{equation}
\begin{array}{c}
(\frac{\partial S}{\partial {\frak M}})_{{\frak J}}=\widehat{\beta }_{H} \\ 
(\frac{\partial S}{\partial {\frak J}})_{{\frak M}}=-\widehat{\beta }_{H}%
\widehat{\Omega }_{H}%
\end{array}
\label{1law}
\end{equation}%
are satisfied with our result obtained in equations (\ref{KdSangvel}), (\ref%
{betaH}), (\ref{totangumom}) and (\ref{KdSoutent2}) where $\widehat{\beta }%
_{H}=\frac{\beta _{H}}{\Xi }$ and $\widehat{\Omega }_{H}=\Omega _{H}\Xi -%
\frac{a}{\ell ^{2}}.$ So we find that the first law of thermodynamics 
\begin{equation}
d{\frak M=}\frac{1}{\widehat{\beta }_{H}}dS+\widehat{\Omega }_{H}d{\frak J}
\label{dm}
\end{equation}%
is valid, where $\widehat{\Omega }_{H}$ $=\frac{a(\ell ^{2}-\tau _{c}^{2})}{%
\ell ^{2}(\tau _{c}^{2}+a^{2})}$ is the angular velocity of the horizon.

Since the normal at past/future infinity is timelike and not spacelike some
care must be taken in physically interpreting this quantity. \ For the Kerr
and Kerr-AdS spacetimes, the closest notion one has to a family of static
observers outside of a black hole is that of locally non-rotating observers,
whose coordinate angular velocity is given by $\Omega =-\frac{g_{t\phi }}{%
g_{\phi \phi }}$. In a stationary spacetime this will be a function of
radial position, diminishing to either zero (for Kerr) or a finite value
(for Kerr-AdS), which corresponds to an observer's angular velocity at
infinity. \ However the situation for Kerr-dS is that of a collection of
observers outside of a cosmological horizon, where $\partial /\partial t$\
is spacelike. The notion of angular velocity becomes that of a helical path
that a given observer traces as he/she moves along curves at future (past)
infinity whose tangent vectors are $\partial /\partial t$. \ In this case
the \ $\Omega =-\frac{g_{t\phi }}{g_{\phi \phi }}\longrightarrow \Omega
_{\infty }=\frac{a}{\ell ^{2}+a^{2}}$\ for large $\tau $; this corresponds
to the rate of change of the angular position of the frame of reference with
respect to the $t$-direction. \ Similarly, the angular momentum is conserved
from place to place anywhere along curves at future (past) infinity whose
tangent vectors are $\partial /\partial t$: \ a collection of observers on a
constant-$t$\ hypersurface (which itself does not enclose any bulk space %
\cite{Ivannonortho}) all observe the same value of ${\frak J}$ regardless of
the value of $\ t$. \ We shall continue to use the terms ``angular
velocity'', ``angular momentum'', and ``rotation at infinity'' keeping in
mind this distinction in physical interpretation relative to the
asymptotically flat and AdS cases.

The quantity $\widehat{\Omega }_{H}$\ is measured with respect to a frame
that is not rotating at infinity, and differs from $\Omega _{H}$\ , which is
the corresponding quantity with respect to a frame that is rotating at
infinity with angular velocity $\Omega _{\infty }$. It is the angular
velocity $\widehat{\Omega }_{H}$\ that appears in the first law, and is
related to $\Omega _{H}$\ by{\bf \ }$\widehat{\Omega }_{H}/\Xi =\Omega
_{H}-\Omega _{\infty }$. The satisfaction of the first law of thermodynamic
for Kerr-dS spacetimes with given conserved mass (\ref{totmass}), angular
momentum (\ref{totangumom}) and entropy (\ref{KdSoutent2}) guarantees that
the ratio of the entropy to the horizon area has the expected value of $1/4.$%
\ The importance of using a frame that is not rotating at infinity has
previously been been noted for Kerr-AdS black holes \cite{Hossein}, and a
discussion of its relevance for the first law of thermodynamics has recently
appeared \cite{Gidd}. What is interesting here is that the same distinction
is required, despite the difference in physical interpretation in the
asymptotically dS case.

\begin{figure}[tbp]
\begin{center}
\epsfig{file=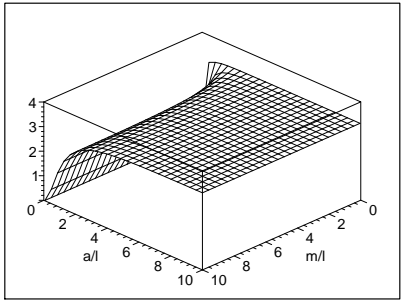,width=0.3\linewidth}
\end{center}
\caption{Entropy of Kerr-dS spacetime with respect to rescaled mass $m/\ell $
and rotational parameter $a/\ell $. The horizontal sheet denotes the N-bound
of $\protect\pi \ell ^{2}$. }
\label{fig1}
\end{figure}

Figure (\ref{fig1}) shows the typical behavior of $S$ as a function of
rescaled mass $m/\ell $ and rotational parameter $a/\ell $. \ We note that
when $a=0,$ the cosmological horizon is located at $\tau _{c}=0$ for all
values of $m$ and the entropy is $S=0$. By increasing the rotational
parameter for a fixed $m$, the entropy monotonically increases and
approaches the N-bound. As is clear from figure (\ref{fig1}) in the special
case of $m=0$, for all values of $a$ the entropy is equal to the N-bound.
This is as expected since the $m=0$ metric is just pure dS spacetime in
unusual coordinates. Note also the behavior of the entropy as the black hole
approaches extremality. The mass and rotational parameters of the extremal
black hole are given by%
\begin{equation}
\begin{array}{c}
m_{ext.}=\frac{4(2-\sqrt{3})}{3}\ell \sqrt{2\sqrt{3}-3} \\ 
a_{ext.}=(2-\sqrt{3})\ell%
\end{array}
\label{maext}
\end{equation}%
and in this case, the inner horizon and the outer horizon coincide with the
cosmological horizon that is located at 
\begin{equation}
\tau _{c_{(ext.)}}=\frac{\sqrt{6\sqrt{3}-9}}{3}\ell
\end{equation}%
and the fourth root of the $\Delta (\tau )$ (which is a negative and
non-physical root) is $-3\tau _{c_{(ext.)}}.$ Applying the relation (\ref%
{KdSoutent2}), shows that the entropy has a definite value of $\pi \frac{9-5%
\sqrt{3}}{6(2-\sqrt{3})}\ell ^{2}$ satisfying the N-bound. \ 

In the other extremal case, where only the inner and outer horizons
coincide, the cosmological horizon is given by $\tau _{c_{(ext)}}=-\tau
_{in_{(ext)}}+\frac{\ell ^{2}-\tau _{in_{(ext)}}^{2}}{\sqrt{\ell ^{2}+\tau
_{in_{(ext)}}^{2}}}$ and the entropy always respects the N-bound for all the
allowed values of the extremal inner horizon.

Consequently we confirm the expected result that the N-bound on entropy is
satisfied for all the values of the mass and rotational parameter.

Figure (\ref{fig1}) has several interesting features. For fixed $m$\ the
entropy increases for increasing $a$, rapidly approaching the N-bound. For
any value of $a/\ell >2$, the entropy remains very close to the N-bound; in
these situations $m/a$\ is relatively small, and the spacetime very close to
pure dS. Although difficult to see from figure (\ref{fig1}), the upper sheet
slopes slightly downward toward the left of the diagram, so that by
increasing the mass parameter $m,$\ for a fixed $a$, the entropy decreases
and so departs more from the N-bound.{\Large \ }Moreover, we see from eqs.(%
\ref{KdSoutfuncs}) and (\ref{KdSoutent2}) that in the limit $m\rightarrow 0$%
,\ the entropy saturates the N-bound for all values of rotational parameter $%
a$.\ This is as expected since in the limit $m\rightarrow 0$,\ \ the Kerr-dS
metric (\ref{KdSoutmetr}) reduces to the standard form of the dS spactime
under a coordinate transformation.

\bigskip

In both the AdS and dS cases there is a natural correspondence between
phenomena occurring near the boundary (or in the deep interior) of either
spacetime and UV (IR) physics in the dual CFT. Solutions that are
asymptotically (locally) dS lead to an interpretation in terms of
renormalization group flows and an associated generalized dS $c$-theorem.
This theorem states that in a contracting patch of dS spacetime, the
renormalization group flows toward the infrared and in an expanding
spacetime, it flows toward the ultraviolet. In reference \cite{Leb}, a $c$%
-function was defined for a representation of the dS metric with a wide
variety of boundary geometries involving direct products of flat space, the
sphere and hyperbolic space. The definition of $c$-function in these cases,
is based on its generalization from the flat boundary geometry. For a
four-dimensional dS spacetime, it is given by%
\begin{equation}
c=\left( G_{\mu \nu }n^{\mu }n^{\nu }\right) ^{-1}  \label{cfun}
\end{equation}%
where $n^{\mu }$\ is the unit vector in the $\tau $ direction. Consequently
the $c$-theorem implies that the $c$-function (or $\overline{c}$-function)
must increase (decrease) for any expanding (contracting) patch of the
spacetime.

\begin{figure}[tbp]
\begin{center}
{\bf {\Large \epsfig{file=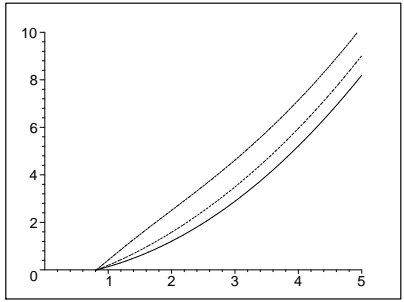,width=0.3\linewidth} } }
\end{center}
\caption{The $c$-function of Kerr-dS spacetime plotted as a function of $%
\protect\tau $ for three different values of angular variable $\protect%
\theta =0$ (lower solid curve)$,\frac{\protect\pi }{5}$ (dotted curve)$,%
\frac{\protect\pi }{3}$ (upper dashed curve). The cosmological constant $%
\Lambda =3$ and the mass and rotational parameters are respectively set to $%
2 $ and $3$. The cosmological horizon is located at $\protect\tau %
_{c}=0.6449.$ }
\label{fig1.65}
\end{figure}

\begin{figure}[tbp]
\begin{center}
\epsfig{file=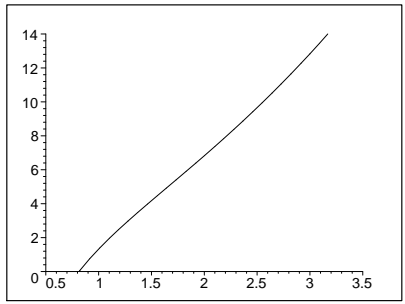,width=0.3\linewidth}
\end{center}
\caption{The $\overline{c}$-function of Kerr-dS spacetime plotted as a
function of $\protect\tau $. The cosmological constant $\Lambda =3$ and the
mass and rotational parameters are respectively set to $2$ and $3$. The
cosmological horizon is located at $\protect\tau _{c}=0.6449.$ }
\label{fig1.7}
\end{figure}

Since the spacetime (\ref{KdSoutmetr}) is asymptotically locally dS, if we
use the relation (\ref{cfun}), we find%
\begin{equation}
c(\tau ,\theta )=\frac{1}{3}\frac{\tau ^{4}+\tau ^{2}(a^{2}-\ell
^{2})+2m\ell ^{2}\tau -a^{2}\ell ^{2}}{\tau ^{2}+a^{2}\cos ^{2}\theta }
\label{cfunKdS}
\end{equation}%
We see that, due to the presence of rotation, the $c$-function depends
explicitly to the angular coordinate $\theta .$ In the language of paper %
\cite{Leb}, this is due to the dependence of the effective cosmological
constant on the angular variable. In figure (\ref{fig1.65}), the $c$%
-functions for three different fixed angular coordinates, are plotted. 
We consider the following quantity%
\begin{equation}
\overline{c}(\tau )=\frac{1}{{\frak A}}\int_{{\frak S}}c(\tau ,\theta )dS
\label{cfuneffKdS}
\end{equation}%
which integration is over the 2-surface ${\frak S}$ (with area ${\frak A}$),
parametrized by $(\theta ,\phi )$ for a fixed $t$ and $\tau $, describing
the average $c$-function of the spacetime (\ref{KdSoutmetr}). Although the $%
c $-function (and RG flow) is anisotropic, as the boundary is approached
this anisotropy vanishes and the $c$-function (or $\overline{c}$-function)
behaves $\sim $ $\tau ^{2}$ in good agreement with one expects from the $c$%
-theorem. In figure (\ref{fig1.7}), the $\overline{c}$-function outside the
cosmological horizon, is plotted. In this plot, we set $\ell =1,m=2$ and $%
a=3.$ The cosmological horizon is located at $\tau _{c}=0.6449.$

\QTP{
}
As we see from the figure, outside the cosmological horizon for any value of 
$\theta $, the $c$-function is a monotonically increasing function of $\tau $%
, indicative of the expansion of a constant $\tau $-surface of the metric (%
\ref{KdSoutmetr}) outside of the cosmological horizon. The metric (\ref%
{KdSoutmetr}) at future infinity ($\tau \rightarrow +\infty $) reduces to 
\begin{equation}
-d\varkappa ^{2}+e^{2\varkappa /\ell }d\digamma _{3}^{2}
\label{KDSfutureinf}
\end{equation}%
where $\chi =\ell \ln \tau $. Here $d\digamma _{3}^{2}$ is the metric of
three-dimensional constant $\chi $-surface which is neither maximally
symmetric nor uniform in the $\theta $-direction. We note that the scale
factor $e^{2\varkappa /\ell }$ in (\ref{KDSfutureinf}) expands exponentially
near future infinity, so the behavior of $\ \overline{c}$-function in figure
(\ref{fig1.7}) is in good agreement with what one expects from the $c$%
-theorem.

\QTP{
}
Although it seems that the $c$-function given in (\ref{cfunKdS})\ is
anisortropic in the limiting case of $m\rightarrow 0,$\ this is not the
case. In fact in the limit $m\rightarrow 0,$\ by a coordinate
transformations, the Kerr-dS metric (\ref{KdSoutmetr}) reduces to the
following standard form of the pure dS spacetime%
\begin{equation}
-(1-\frac{y^{2}}{\ell ^{2}})dT^{2}+\frac{dy^{2}}{1-\frac{y^{2}}{\ell ^{2}}}%
+y^{2}(d\Theta ^{2}+\sin ^{2}\Theta d\psi ^{2})  \label{y}
\end{equation}%
for which the $c$-function is equal to the isotropic value $\frac{y^{2}-\ell
^{2}}{3}$. \ The angular dependence of the $c$-function in (\ref{cfunKdS})\
is a consequence of the anisotropy of the metric functions\ in the
coordinates we are using. \ 

\section{\bf Conclusions}

We have shown that the entropy and the mass of the class of Kerr-dS
spacetimes always respect the N-bound and dS-maximal mass conjectures. These
results hold even when the rotating hole approaches extremality. The first
law of thermodynamics is also obeyed, albeit with differing interpretations
of the physical quantities relative to their asymptotically flat and AdS
counterparts.

For the dS spacetimes with rotation, we have shown that the notion of the $c$%
-function must be extended, describing an anisotropic renormalization group
flow that becomes completely isotropic as the boundary is approached. The
behavior of the averaged $c$-function is in complete agreement with the $c$%
-theorem.

Our results also provide substantive evidence against any stronger
formulation of the maximal mass conjecture. Consider a Kerr-dS spacetime for
sufficiently small mass and zero rotation. In this case there is a
cosmological horizon, a negative mass at future infinity and an outer
horizon censoring the singularity, in accord with the stronger version of
the conjecture. For small nonzero rotation this situation is unchanged. But
as the rotation parameter grows, the inner \& outer horizons get closer
together and eventually they coincide. For larger values of the rotation
parameter, there is a naked singularity. However the mass at future infinity
is still negative, ie still less that dS spacetime. For example, a spacetime
with $m=\ell /10$\ and $a=\ell /12$\ does not have a cosmological (naked)
singularities, whereas a spacetime with $m=\ell /10$\ and $a=\ell /4$\ does.
\ Yet both have negative mass at future infinity. In general the spacetime
will have a naked singularity with negative mass at future infinity whenever 
\begin{equation}
\widehat{m}<\frac{\sqrt{6-6\widehat{a}^{2}-6\sqrt{(\widehat{a}^{2}-4\widehat{%
a}+1)(\widehat{a}^{2}+4\widehat{a}+1)}}}{18}\left( 2\left( 1-\widehat{a}%
^{2}\right) +\sqrt{(\widehat{a}^{2}-4\widehat{a}+1)(\widehat{a}^{2}+4%
\widehat{a}+1)}\right)  \label{m}
\end{equation}%
where \ $\widehat{a}=\frac{a}{\ell }$\ and $\widehat{m}=\frac{m}{\ell }$.
Consequently there is a class of Kerr-dS spacetimes that have singularities,
but still satisfy the (weaker) maximal mass conjecture. \ It is difficult to
see how the conjecture could be strengthened in a manner that \ would render
it useful for a significantly broad class of physically interesting
spacetimes.

We conclude with a few comments about directions for future work. An obvious
thing to consider are higher dimensional Kerr-dS spacetimes with multiple
rotation parameters \cite{PAge}; the behavior of the $c$-function in this
case should exhibit a significantly greater degree of anisotropy. Another
avenue to explore is the validity of the entropy-area relation $S=A/4$,
which is satisfied for any black hole in a $(d+1)$ dimensional
asymptotically flat/AdS setting, where $A$ is the area of a $d-1$
dimensional fixed point set of isometry group. However, entropy can defined
for other kinds of spacetimes in which the isometry group has fixed points
on surfaces of even co-dimension \cite{Haw}. The best examples of these
spacetimes are asymptotically locally flat and asymptotically locally anti
dS spacetimes with NUT charge. In these cases when the isometry group has a
two-dimensional fixed set (bolt), the entropy of the spacetime is not given
by the area-entropy relation, as a consequence of the first law of
thermodynamics \cite{as}.

In asymptotically dS spacetimes, the Gibbs-Duhem entropy is proportional to
the area of the horizon and respects the N-bound (for the other case of \
Schwarzschild-dS spacetime, see \cite{GM}). However, for asymptotically
locally dS spacetime with NUT charge, the entropy is no longer proportional
to the area. Consequently the entropy need not respect the N-bound, an
expectation confirmed by investigations for a wide range of situations with
nonzero NUT charge \cite{CGM3, Anderson}. We are currently considering the
validity of the N-bound and maximal mass conjectures for NUT charged
asymptotically de Sitter spacetimes with rotation, and plan to report our
results in a future publication \cite{GM2}.

\bigskip

{\Large Acknowledgments}

This work was supported by the Natural Sciences and Engineering Research
Council of Canada. A.M.G. would like to thank R. Myers for helpful
discussions.

\ 


\bigskip

\begin{thebibliography}{99}
\bibitem{bala} V. Balasubramanian, J. de Boer and D. Minic, {\it Phys. Rev.} 
{\bf D65}, 123508 (2002).

\bibitem{balakraus} V. Balasubramanian and P. Kraus, {\it Commun. Math. Phys.%
} {\bf 208}, 413 (1999).

\bibitem{brown} J.D. Brown and \ J.W. York, {\it Phys. Rev.} {\bf D47}, 1407
(1993).

\bibitem{BCM} J.D. Brown, J. Creighton and R.B. Mann, {\it Phys. Rev.} {\bf %
D50}, 6394 (1994).\ 

\bibitem{ivan} I.S. Booth and R.B. Mann, {\it Phys. Rev.} {\bf D59}, 064021
(1999).

\bibitem{Klem} D. Klemm, {\it Nucl. Phys.} {\bf B625}, 295\ (2002).

\bibitem{Myung} Y.S. Myung, {\it Mod. Phys. Lett.} {\bf A16}, 2353 (2001).

\bibitem{GM} M. Ghezelbash and R.B. Mann, {\it JHEP} {\bf 0201}, 005 (2002).

\bibitem{Dumitru} D. Astefanesei, R. Mann, E. Rau, {\it JHEP} {\bf 0401},%
{\bf \ }029 (2004).

\bibitem{cai} R.G. Cai, Y.S. Myung and Y.Z. Zhang, {\it Phys. Rev.} {\bf D65}%
, 084019\ (2002).

\bibitem{bousso} R. Bousso, {\it JHEP} {\bf 9907}, 004 (1999);\ {\it JHEP} 
{\bf 0011}, 038 (2000).

\bibitem{CGM1} R. Clarkson, A.M. Ghezelbash and R.B. Mann,\ {\it Phys. Rev.} 
{\it Lett.} {\bf 91}, 061301 (2003).

\bibitem{CGM2} R. Clarkson, A.M. Ghezelbash and R.B. Mann, {\it Nucl. Phys.} 
{\bf B674}, 329 (2003).

\bibitem{stro} A. Strominger, {\it JHEP} {\bf 0111}, 049 (2001).

\bibitem{GM2} M. Ghezelbash and R.B. Mann, {\it in preparation.}

\bibitem{ivanrobb} I.S. Booth and R.B. Mann, {\it Nucl. Phys.} {\bf B539},
267\ (1999); I.S. Booth and R.B. Mann, \ {\it Phys.Rev.Lett.} {\bf 81}, 5052
(1998).

\bibitem{simon} R.B. Mann and S.F. Ross, {\it Phys. Rev.} {\bf D52}, 2254
(1995).

\bibitem{WALD} R.M. Wald, ''General Relativity'', University of Chicago
Press, Chicago, 1984.

\bibitem{Deh} M.H. Dehghani and H. KhajehAzad, {\it Can. J. Phys.} {\bf 81},
1363 (2003).

\bibitem{Ivannonortho} I.S. Booth and R.B. Mann, {\it Phys. Rev.} {\bf D59},
064021 (1999).

\bibitem{Hossein} M.H. Dehghani and R.B. Mann, {\it Phys.Rev.} {\bf D64},\
044003 (2001); V.A. Kostelecky and M. J. Perry, {\it Phys. Lett.} {\bf B371, 
}191 (1996).

\bibitem{Gidd} G.W. Gibbons, M.J. Perry and C.N. Pope, {\it hep-th}/0408217.

\bibitem{Leb} F. Leblond, D. Marolf and R.C. Myers, {\it JHEP} {\bf 0206},
052 (2002).

\bibitem{PAge} G.W. Gibbons, H. Lu, D.N. Page and C.N. Pope, {\it Phys. Rev.
Lett.} {\bf 93}, 171102 (2004).

\bibitem{Haw} S.W. Hawking and C.J. Hunter, {\it Phys. Rev.} {\bf D59},
044025 (1999).

\bibitem{as} D. Astefanesei, R.B. Mann and E. Radu, {\it hep-th/}0406050.

\bibitem{CGM3} R. Clarkson, A.M. Ghezelbash and R.B. Mann, {\it Int. J. Mod.
Phys.} {\bf A19}, 3987 (2004).

\bibitem{Anderson} Micheal T. Anderson, {\it hep-th}/0407087.
\end{thebibliography}
\end{document}